\documentclass[12pt,preprint]{aastex}
\newcommand{\totalnum}{72}
\newcommand{\ldnum}{65}
\newcommand{\sdnum}{7}
\newcommand{\opticalagnum}{47} 
\newcommand{\Xoverlapnum}{14}
\newcommand{\Xoverlapexp}{6.7}
\newcommand{\Xmontecarlo}{0.5\%} 
\newcommand{\Xmedianprob}{6.9\%}

\newcommand{\Emedianarea}{1.6\%}
\newcommand{\EmedianareaOAG}{1.6\%} 
\newcommand{\Eexpected}{1.3} 
\newcommand{\EexpectedOAG}{0.82}

\newcommand{\Emedianareawcut}{1.4\%}
\newcommand{\EmedianareawcutOAG}{1.3\%}
\newcommand{\Eexpectedwcut}{1.2}
\newcommand{\EexpectedwcutOAG}{0.73}

\newcommand{\Imedianarea}{1.0\%} 
\newcommand{\ImedianOAG}{1.0\%}
\newcommand{\Iexpected}{0.86}
\newcommand{\IexpectedOAG}{0.54}

\newcommand{\Imedianareawcut}{0.9\%}
\newcommand{\ImedianOAGwcut}{0.9\%}
\newcommand{\Iexpectedwcut}{0.79} 
\newcommand{\IexpectedOAGwcut}{0.50}

\shorttitle{GRBs and Host Galaxies}
\shortauthors{Cobb & Bailyn}

\begin{document}

\title{Connecting GRBs and galaxies: the probability of chance coincidence\altaffilmark{1}}

\author{Bethany E. Cobb\altaffilmark{2} and Charles D. Bailyn\altaffilmark{2}}
\email{cobb@astro.yale.edu}

\altaffiltext{1}{Using observations from the Small and Moderate Aperture Research Telescope System (SMARTS) consortium, which operates 4 telescopes at the
   Cerro Tololo Inter-American Observatory}

\altaffiltext{2}{Department of Astronomy, Yale University, P.O. Box 208101, New Haven, CT 06520}

\begin{abstract}
Studies of GRB host galaxies are crucial to understanding GRBs.
However, since they are identified by the superposition
in the plane of the sky of a GRB afterglow and a galaxy there is always a
possibility that an association represents a chance
alignment, rather than a physical connection.  
We examine a uniform sample of \totalnum\ GRB fields to explore
the probability of chance superpositions.  There is
typically a $\sim1$\% chance that an optical afterglow
will coincide with a galaxy by chance.
While spurious host galaxy detections will, therefore, be rare, the possibility must be
considered when examining individual GRB/host galaxy examples.
It is also tempting to use the large and uniform collection of X-ray afterglow positions
to search for GRB-associated galaxies.  However, we find
that approximately half of the 14 superpositions in our sample are likely to
occur by chance, so in the case of GRBs localized only by an X-ray afterglow, even
statistical studies are suspect.
\end{abstract}


\section{Introduction}
From the earliest associations between gamma-ray bursts (GRBs) and host galaxies, burst environments have revealed many
important clues to the nature of GRB progenitors \citep[e.g.][]{Sokolov+01,Bloom+02,Fynbo+03,LeFloch+03}.  As \textit{Swift} 
\citep{Gehrels+04} accurately localizes
hundreds of GRBs, statistical studies of GRB host galaxies are becoming increasingly critical to
understanding GRB formation mechanisms \citep[e.g.][]{Christensen+04,Wainwright+07}. 

A GRB host galaxy is identified by the superposition in the plane of the sky of the GRB afterglow (AG) and a galaxy.
With only visual associations between afterglows and galaxies used to identify GRB host galaxies, the possibility always
exists for an incorrect association.
Some afterglow and galaxy associations may be chance superpositions, with the galaxy either a
foreground or background galaxy that is not physically associated with the GRB.
These mistaken associations could cause confusion when analyzing data on GRB environments.
When dealing with a statistical sample of GRBs, rather than a single case-study, results will only
be impacted if a relatively significant number of the host galaxies have been mistakenly identified.
Understanding the likelihood of galaxy mis-identification in a given sample of GRBs is, therefore, imperative when
conducting such studies.
However, even a single mistaken association might generate confusion if the combination of
GRB and galaxy characteristics are anomalous.
For example, the potentially paradigm-shifting object GRB 060614 \citep[e.g.][]{Cobb+06,DellaValle+06,Gal-Yam+06,SX06},
a burst identified as low-redshift for which no SN was detected, could
be a typical object of no special interest if the burst's purported host galaxy is actually just a random galaxy
along the line-of-sight to the GRB.

When an optical afterglow is detected, afterglow localization is precise to significantly less than an arcsecond.  
The superposition between a galaxy and an optical afterglow is, therefore,
generally taken as proof that the galaxy in question is the host galaxy of the GRB.  
Since the optical afterglow position can be determined to within sub-pixel accuracy,
the probability of the optical afterglow falling randomly on any galaxy-covered pixel
is determined by dividing the number of pixels covered by galaxies in each field by
the total number of pixels in that field.
If only an X-ray afterglow is detected, the GRB can only be localized to within an error region of radius $\sim2$ arcseconds.  
If a galaxy is detected within this error
region it is generally assumed to be the host galaxy of the GRB, although confusion can arise when two or more sources
are detected within or around a single X-ray afterglow error region \citep[e.g.][]{Ferrero+07}.  

In this paper we investigate a uniform set of \totalnum\ GRB fields in order to examine the general
issue of associations between GRBs and host galaxies.  In \S 2, we describe our data and analysis techniques.
In \S 3, the probability of chance coincidence between optical afterglows
and galaxies is determined by measuring the fractional area covered by galaxies.  
In \S 4, localizations based on X-ray afterglows are considered by measuring the probability
of galaxies falling within randomly placed X-ray error regions. Afterglow-associated galaxies are compared
to field galaxies in \S 5.  We discuss our results in \S 6 along with some
strategies for recognizing false hosts and conclude in \S 7.

\section{Data}
Our I-band optical images were obtained using the ANDICAM instrument mounted on the 1.3m telescope at Cerro Tololo Inter-American 
Observatory.\footnote{http://www.astronomy.ohio-state.edu/ANDICAM}
This telescope is operated as part of the Small and Moderate Aperture Research
Telescope System (SMARTS) consortium.\footnote{http://www.astro.yale.edu/smarts} 

\subsection{GRB Selection}
The \totalnum\ bursts included in these analyses were, in general, selected based on their observability from CTIO, limited
to those bursts with declination $\lesssim +35\degr$. Observations were also limited by each burst's right 
ascension because all observations began within days post-burst.
Bursts were occasionally not observed due to telescope scheduling time limitations.  
Both long-soft bursts (\ldnum) and short-hard bursts (\sdnum) were observed. 
Due to large and uncertain values of galactic reddening near the galactic plane, no GRBs with galactic latitude 
$\left|b\right|<10\degr$ are included in this sample.  

The redshifts, or redshift upper limits, of 37 bursts in our sample have been determined through either 
optical afterglow observations or observations of possible host galaxies. 
If only the 21 redshifts that are measured directly from GRB afterglow absorption are considered, then the median redshift is $z=2.3$. 
For the entire sample of bursts, the median redshift is $z=1.3$ with a range from 0.089 to 6.6.  
The median redshift is understandably reduced when including GRB redshifts obtained from associated galaxies, 
because galaxies are generally easier to detect and observe spectrographically when at low redshift.

The GRBs included in this sample are listed in Table 1. 

\subsection{Observations and Data Reduction}
The nightly data set for each burst consisted of 6 individual 360-second I-band observations, taken at slightly offset telescope positions.  
Each individual image was reduced in the typical manner, with bias and dark subtraction, flat fielding,
and cosmic ray removal using the L.A. Cosmic program\footnote{http://www.astro.yale.edu/dokkum/lacosmic/} \citep{vanDokkum01}.
The individual images were median combined with maximum pixel rejection to produce a source-free initial fringe correction image.
This background image was scaled and subtracted from the individual images, which were then aligned and combined.  
This combined image was used to produce a source mask and a new fringe correction image was
produced by recombining the original images -- only this time masking the sources in each image, 
rather than using minmax rejection.  This new background was scaled and subtracted from the individual images, 
which were then aligned and combined to produce the final nightly science frame.
   
For each GRB, between 1 and 11 usable nightly images were obtained (the mode being 4 images).  
Particularly shallow images or nights with relatively poor seeing were excluded.  Some images taken at early times post-burst
were also excluded as they contained optical afterglow light.  All usable nightly images were aligned and combined
to produce a final deep frame for each GRB.  The frame edges were cropped so that each final frame is of
uniform depth over the entire field.  Field size varied slightly but is typically $5\farcm6\times5\farcm4$.
A few fields contained saturated stars with 
diffraction spikes; the area immediately surrounding these stars is excluding from all of the following analyses. 

Secondary standard stars in each field were photometrically calibrated by comparison, on photometric nights,
with Landolt standard stars \citep{Landolt92}.  Photometric calibration is typically
accurate to 0.05 magnitudes.  No photometric observations
were obtained for 4 GRBs and those fields are calibrated using USNO-B1.0 I2
magnitudes \citep{Monet+03}, with typical calibration errors of 0.2 magnitudes.
The reddening corrected \citep{Schlegel+98} $3\sigma$ I-band limiting magnitudes for point sources in these images ranges from
I=21.2 mag to 23.3 mag, with a median of 22.4 mag.  All magnitudes are given in the Vega system.

The fields are astrometrically calibrated using USNO-B1.0 stars, with statistical error of $<0.2"$. The
CCD pixel scale is 0.37"/pixel, so that the astrometry is accurate to within a single pixel.
X-ray afterglow coordinates and 90\% confidence error radii for all bursts are taken from \cite{Butler07}.
Error region radii vary from $0\farcs5$ to $6\farcs9$, with a median of $1\farcs9$.

The data are summarized in Table 1.

\subsection{Galaxy Detection}
All objects in the field were cataloged using SExtractor \citep{BA96} with a 2$\sigma$ detection threshold.
The SExtractor parameter CLASS\_STAR was then used to distinguish the galaxies from the stars in the field.  This parameter
is determined by a neural network and ranges from CLASS\_STAR=0.0 for extended objects to 
CLASS\_STAR=1.0 for point-like objects. We consider objects with CLASS\_STAR$>0.8$
to be stars and these objects are not included in the following analyses.
This value is chosen to avoid contaminating the sample with stars that are mistakenly
classified as galaxies.  Adopting a less strict value results in more galaxy identifications.  
Defining galaxies to have CLASS\_STAR$<0.9$, for example, increases
chance coincidence probabilities by $\sim10$\%. 
In crowded starfields, SExtractor's ability to differentiate between stars and galaxies is reduced.
This effect is minimized in our analyses, however, as none of the fields considered here are within $10\degr$ of the galactic plane.
Galaxy I-band magnitudes are given by MAG\_AUTO, with zeropoints appropriately 
adjusted for each field.  These magnitudes are generally accurate to within a few tenths of a 
magnitude. 

Only galaxies detected by SExtractor are considered in these analyses.  At
the position of a few afterglows there are low significance galaxies
that are not identified by SExtractor. 
For consistency, these visual identifications are not included.

Since our observations are limited to galaxies brighter than $\sim23$ mag, we would not expect
to be able to detect all the potential host galaxies in our images.  
Galaxies go undetected primarily because of their large redshifts, though dwarf galaxies
might be missed even at low redshifts.  
Examination of galaxies brighter than the median limiting magnitude of our sample ($I<22.4$)
in the VVDS-CDFS galaxy catalog \citep{LeFevre+04} suggests that nearly all of the galaxies detected in our
fields are likely to be at a redshift lower than $z=1.5$ and about 65\% have redshift $z\leq0.7$.

To understand our ability to detect GRB host galaxies, it would be useful to know 
the characteristics of a typical GRB host.  In pre-\textit{Swift} GRB samples, however, 
galaxies associated with GRBs have been noted to be a rather
heterogeneous group \citep[e.g.][]{Conselice+05,LeFloch+03} covering a wide range of galaxy types and absolute
magnitudes ($M_B\sim-16$ to $-21$) \citep{Sokolov+01,Conselice+05,Wainwright+07}, 
thought there is a tendency for GRBs to occur in faint galaxies \citep{Jaunsen+03,LeFloch+03,Fruchter+06}.
Assuming a typical galaxy color of $B-I=2$ \citep{Fukugita+95}, the brightest host galaxies could have
have $I=-23$ and might be detected in our images to $z\sim1$.  Observed galaxy brightness would, of course,
depend on the necessary $k$-correction, which depends strongly on galaxy morphology.
The star-forming galaxies of long-duration bursts would be favored for detection at higher redshifts because
of their strong rest-frame UV emission.  Unfortunately, redshift limits for detecting the dimmest host galaxies
are much more severe. These dwarf hosts may be common in the local universe, but many could not
be detected in our images to even moderate redshifts given our magnitude limits.

\section{Optical Afterglow and Galaxy Coincidence}
There is no exact projected distance at which a GRB and a galaxy become associated,
and galaxies lack clear ``edges''. Therefore, we calculate the probability
of a chance optical afterglow/galaxy association below using two slightly different definitions of 
galaxy area.

\subsection{ISOAREAF\_IMAGE area}
One measure of the pixel area of each galaxy is given by the isophotal SExtractor parameter ISOAREAF\_IMAGE.
This value disregards the lower significance outskirts of each galaxy, so it is a conservative estimate of galaxy area.
If an optical afterglow were to fall within the ISOAREAF\_IMAGE area of a galaxy,
that galaxy would be regarded as a strong host-galaxy candidate.
This excludes cases in which GRBs fall at the outskirts of galaxies -- which are bound to occur.
Hence these figures serve strictly as a lower limit.

The ISOAREAF\_IMAGE area covered by galaxies in each field ranges from 0.2\% to 3.5\% (see Figure 1 and Table 1), with a median of \Imedianarea\ 
(see Table 2).
Summing the overlap probability over all \totalnum\ fields yields an expectation of \Iexpected\ observed galaxy/optical afterglow
coincidences -- including even those fields for which no optical afterglow detection was made.
For the \opticalagnum\ fields in which optical afterglow was detected from either short- or long-duration bursts,
there is an expectation of \IexpectedOAG\ observed galaxy/optical afterglow coincidences.
In fact, this sample contains 7 such coincidences (GRBs 050416a, 050724, 050826, 060505, 060614, 061021 and 061121).
Based on a Monte Carlo simulations, the chance of having randomly observed that many coincidences is less than 0.004\%.
These results are summarized in Table 2.

Since the total galaxy-covered area in each field depends on limiting magnitude, which varies from field to field, it
is useful to consider a more homogeneous selection of galaxies.  As deeper observations
are obtained, more galaxies will be detected.  Therefore, deeper imaging increases
the chance for a correct identification, but also increases the chance 
for spurious associations because the fractional area covered by galaxies is increased.
By imposing a magnitude cutoff, therefore, we produce
a minimum value for the possibility of chance superposition.
For the homogeneous sample, we limit our analysis to galaxies with Galactic extinction corrected magnitudes brighter than $I=21.5$.
This value is determined by producing a magnitude histogram of all the galaxies being used in this analysis, with bin
size of half a magnitude, and selecting the midvalue of the bin containing the maximum number of galaxies (see Figure 2).
Beyond this magnitude, our galaxy sample is significantly affected by 
incompleteness.  In individual fields, this galaxy completeness ``turnover" value ranges from 20 to 22, with
a median and mode of 21.5. In the fields that are incomplete at 21.5 mag, there will be ``missing'' galaxies.  
This means that the probability of chance superposition with a $I\leq21.5$ galaxy will be slightly underestimated.

Not surprisingly, the exclusion of all $I>21.5$ galaxies reduces the field coverage of galaxies by only a small amount (see
Table 2).  The most significant change is that only 4 of the 7 observed optical afterglow/galaxy associations in this sample
occur with galaxies brighter than 21.5 magnitudes (GRBs 050724, 050826, 060505, 061021).  However, Monte Carlo
simulations indicate that the chance of having observed these 4 at random in the \opticalagnum\ fields in 
which optical afterglow was detected is $<1$\%.
So even in this reduced sample, there remains a meaningful overdensity of galaxy/optical afterglow coincidences.
However, it is plausible that one or more of these associations could be a coincidence.

\subsection{Ellipse Area}
Without fully understanding how all GRBs are produced, the exact placement of GRBs within their galaxies cannot
be accurately predicted.  Long-duration GRBs, for example, may require the kind
of rapid star formation regions often found in galactic spiral arms \citep{Conselice+05,Fruchter+06}, while short-duration GRBs may
favor the fringes of galaxies if they are formed by the mergers of compact remnants \citep{Bloom+06a,Berger+07b}.
We have, therefore, been somewhat too restrictive in assuming that a galaxy will only
be identified as a host if the GRB occurs within the galaxy's ISOAREAF\_IMAGE area.

To consider the more general situation of a optical afterglow at a given observed position
relative to a nearby galaxy, we define galaxy area as an ellipse having major and minor axes of
length $v\times$A\_IMAGE and $v\times$B\_IMAGE, where A\_IMAGE and B\_IMAGE
are SExtractor parameters that represent the maximum and minimum spatial \textit{rms} of each 
galaxy profile and $v$ is a simple scaling factor.  In general,
an ellipse with $v=3$ is visually coincident with the extent of the galaxy (see Figure 3).
When the shape of each galaxy is defined in this way, galaxy area is equal to $v^2\pi\times$A\_IMAGE$\times$B\_IMAGE.

In Figure 4, we plot the probability of chance alignment for an afterglow contained
within a galaxy ellipse with scale factor $v$.  The $v$ values of the 7 GRBs with optical afterglow/galaxy coincidences are marked with arrows.
The probability of chance alignment rises quadratically with $v$ 
until $v$ is so large that the galaxies significantly overlap with one another.  We
plot both the entire sample of galaxies and only those galaxies with $I\leq21.5$.  
Excluding the dimmer galaxies only changes the random overlap probability by a few tenths
of a percent at large $v$. On the right hand side of the graph, the number of expected chance coincidences in our sample
of \totalnum\ GRBs is shown.  

Setting the scale factor to $v=3$, we use the ellipse galaxy area to
repeat the analyzes that were done in \S3.1 with the ISOAREAF\_IMAGE area.  These results are
shown in Table 2.  For individual galaxies the ellipse area is generally larger than the ISOAREAF\_IMAGE area
so using the ellipse area results in a slightly increase probability of chance coincidence.

\subsection{Comparing Short- and Long-Duration GRB Fields}
A comparison of short- and long-duration fields is of interest, although
the sample of short duration bursts in this sample is limited to only \sdnum\ fields (versus
\ldnum\ long-duration bursts).  The galaxy area coverage spans a similar range of values between the two samples.
The median area (ISOAREAF\_IMAGE area) covered by all galaxies in short-GRB fields ($\sim0.009$) is only slightly lower than the median
area covered in long-GRB fields ($\sim0.01$).  Since short bursts may occur further outside
their galaxies than long-duration bursts, the galaxy area over which a true association
would be physically plausible may be larger for short bursts than for long bursts.

In this sample, three short-GRBs had detected optical afterglows, but there is only 
one short-burst galaxy/optical afterglow association.
This association frequency (1 of 3) is somewhat larger than that of the long-GRBs (6 of 44), but this may
be an artifact of the limited short-burst sample. 
While the exact nature of short bursts is not yet clear, if short bursts are associated with early-type galaxies, they 
may be preferentially located in local galaxy clusters \citep[e.g.][]{Pedersen+05,Bloom+06a}.
In that case, a higher frequency of associations might be expected for short bursts than long bursts.
However, the correlation between short bursts and clusters is not yet confirmed. 
The observations presented here are consistent with recent evidence that suggests that not all short bursts 
are limited to local clusters \citep{Berger+07a,Berger+07b}.

\section{X-ray Afterglow and Galaxy Coincidence}
For a non-negligible population of \textit{Swift} GRBs (nearly 50\%), no optical afterglow is detected.
Underluminous optical afterglows and heavy line-of-sight extinction may account for many ``dark'' bursts.
An optical afterglow might also go undetected due to observing constraints, such
as the relative position of the sun or the moon to the burst's coordinates or poor weather conditions
at optimum observing sites.  Regardless of the reason, such a non-detection generally
means that the burst will only by localized to within a few arcseconds by XRT observations of the burst's X-ray afterglow.
XRT observations comprise a dataset that is significantly more uniform than deep, ground-based optical afterglow observations that,
by necessity, are obtained from a large number of different instruments at varying times post-burst.
This homogeneity, combined with the fact that the X-ray localized dataset contains nearly double
the number of optically localized GRBs, makes it tempting to analyze GRB host galaxies based exclusively
on X-ray AG positions.  We, therefore, examine our data to determine how significantly such
an analysis might be impacted by the presence of falsely identified hosts.

Each X-ray afterglow error region of the GRBs in this sample was examined for coincident galaxies,
and \Xoverlapnum\ are found to overlap with one or more SExtracted galaxies.\footnote{The X-ray afterglow
error region of GRB 060505 overlaps with 2 galaxies.}  Any overlap between
the circle defined by the X-ray afterglow error region and a galaxy ellipse with $v=3$ is considered to be a coincidence
(see Figure 3).  
This definition of galaxy area is used instead of the ISOAREAF\_IMAGE area because an
ellipse defines a clear ``edge'' to the galaxy while SExtractor does not output the exact
boundaries of a galaxy.
 
We then investigated the probability that one or more galaxies will fall within any random
region that is the same size as a burst's X-ray afterglow error region.  For each GRB field, 500 random positions were selected
and each position was assigned a region with a radius equivalent to the X-ray afterglow error radius of the corresponding GRB.
All region/galaxy overlaps were then counted.  Only a handful of overlaps occur in
some fields, while in others more than a quarter of all random regions contain a galaxy.
The median overlap probability is \Xmedianprob.  The probability
of one or more galaxies overlapping any region the size of the burst's X-ray afterglow error region
is given in column 8 of Table 1.
The overlap probability in each field strongly depends on both the radius of the random regions
and the galaxy population density.
The burst with the largest X-ray error region is GRB 050412,
with a radius of $6\farcs9$.  In this field, any randomly placed region of that size has a 32\% chance of
overlapping with a galaxy.
Note that the density of the galaxy population is dependent on
image depth, as shallower images will contain fewer dim galaxies.  Chance probabilities will, therefore,
increase with image depth.
Figure 5 shows a histogram of the probability of region/galaxy overlap in the \totalnum\ GRB fields examined.
Arrows on the graph indicate the fields in which the X-ray afterglow error region did coincide
with one or more galaxies.

The expected number of observed overlaps created by chance in this GRB sample is \Xoverlapexp, which is obtained by summing the overlap probability over
all fields.  This is significantly less than the \Xoverlapnum\ actually observed.
Using a Monte Carlo simulation (see Figure 6), the probability of having observed \Xoverlapnum\
overlaps in this sample is found to be only \Xmontecarlo.  
Thus, there is a clear over-density of galaxies in GRB X-ray afterglow error regions.  
But the galaxies associated with the X-ray afterglow error regions do not have to be the true host galaxies of the GRBs.
In fact, of the X-ray afterglow error regions in this sample that contain a galaxy, it is likely that
approximately half contain a galaxy that is not associated with the GRB.
This does not appear to be an artifact of images in our sample with large X-ray afterglow error region radii. 
When we consider only fields with small error radii, $\leq2"$, we observed 5 overlaps when only 2 are expected.

Approximately 250 \textit{Swift} bursts have been detected and well-localized by their X-ray afterglows using the XRT.  Based simply
on the median probability of overlap in these fields, there is only a $10^{-8}$ chance that \textit{no} ``false hosts'' have been detected.
In fact, if our sample represents a typical distribution then there could be over 20 such detections
in the entire \textit{Swift} sample.  Improving the limiting magnitudes in each field would only serve to significantly
increase the number of galaxy detections.  Indeed, as image depth increases
the detection of one or several galaxies within a given X-ray afterglow error region is inevitable.
Clearly, caution is required when identifying a galaxy in a X-ray afterglow error region
as the host galaxy of the GRB.  Note, in particular, that relatively few optical afterglow have been 
associated with short-duration bursts, yet many claims as to the nature of short-duration 
bursts have recently been made on the basis of their X-ray afterglow and galaxy associations \citep[e.g.][]{Pedersen+05,Bloom+06a}.
The likelihood of misidentified hosts is, therefore, of particular concern when considering short bursts.

\section{Field Galaxies versus Afterglow-Associated Galaxies}
Having collected the observable properties of all the galaxies in these \totalnum\ fields, we
can determine if the galaxies associated with either X-ray or optical afterglows represent
an anomalous sample of the field galaxies.  The observable properties considered are magnitude, pixel area (either ISOAREAF\_IMAGE
area or ellipse area) and ellipticity ($e$), where $e\equiv1$-(B\_IMAGE/A\_IMAGE).  These observable properties
are, of course, a function of intrinsic size, shape, brightness and redshift.  

Figure 7a shows galaxy ellipse area, with $v=3$, versus magnitude.  Monte Carlo simulations show that
the galaxies associated with X-ray afterglow error regions are a typical sample of all
the galaxies in the field over this dimension.  This result is not surprising, since nearly
half these galaxies could actually be just field galaxies, rather than GRB-associated galaxies.
The galaxies associated with optical afterglow, however, do appear slightly different than the field
galaxies at the $\sim2\sigma$ level.  This seems to be due to the fact that these galaxies
are somewhat brighter than typical field galaxies: $\sim30$\% of optical-afterglow-associated galaxies
have $I<18$ mag, while this is true of only $\sim3$\% of field galaxies.  
To test this, we split the entire sample of galaxies into three magnitude bins:
bright: $I<18$, intermediate: $18\leq I<22$ and faint: $I\geq22$.  Of the 7 galaxies associated with optical afterglow, 
2 are bright, 3 are intermediate and 2 are faint.  We then repeat the Monte Carlo
simulation but require that for each run the 7 galaxies randomly picked from the entire
sample include 2 bright, 3 intermediate and 2 faint galaxies.
With the addition of this magnitude selection, the optical afterglow-associated galaxy sample
becomes indistinguishable from the field galaxies.

Similar analyses are performed for galaxy ISOAREAF\_IMAGE area versus magnitude,
ellipticity versus magnitude, galaxy ellipse area versus ellipticity and
galaxy ISOAREAF\_IMAGE area versus ellipticity (see Figure 7b-e).  The
X-ray afterglow error regions are a typical sample of all the galaxies in the field over these dimensions. 
The optical afterglow-associated galaxies are atypical on the $\sim2\sigma$ level when comparing
ISOAREAF\_IMAGE area versus magnitude and ellipticity versus magnitude. Again, this seems to 
be due to the fact that these galaxies are somewhat brighter than the average field galaxies.

\section{Discussion}
Unlike the fairly uniform images considered here, most observations of GRB fields are obtained
with a wide range of depth and resolution.  Improving depth and resolution
will result in the detection of more galaxies, in which case the values calculated
here serve only as a lower limit to the probability of chance association.  
While increased resolution may seem to make the ``edge'' of a galaxy more defined,
we would caution that knowing the exact optical extent of the galaxy
is not helpful for identifying a GRB's host without first fully understanding
how GRBs trace light.

An additional complication is lensing, which will always occur
when a background GRB is observed through a foreground galaxy.  
Lensing will increase the brightness of the GRB and its afterglow
and shift the observed position of the GRB's afterglow relative to the lensing
galaxy \citep{Paczynski86,Mao92,Grossman+Nowak94}.  
The magnification produced by this lensing, of course, depends strongly on galaxy mass,
observer to lens distance, lens to source distance, and the extent of the alignment.
This lensing effect actually increases the likelihood of detecting a chance association
because it increases the observed brightness of a GRB's optical afterglow.
Host galaxy/optical afterglow associations, by definition, require the detection of an optical afterglow,
and brighter optical afterglows are more likely to be detected than dim afterglows.

If pseudo-redshifts derived from GRB luminosity relationships eventually become accurate enough to use
in evaluating afterglow/galaxy coincidences, lensing would present a complication because lensed
GRBs might not follow expected luminosity relationships.
The Amati relationship \citep{Amati+02,Amati06}, for example, correlates gamma-ray spectral peak energy with a burst's isotropic equivalent
radiated energy.  From measurements of peak spectral energy and gamma-ray flux, therefore, a
GRB's redshift can be estimated. Lensing should not alter the gamma-ray burst spectral shape so that
peak energy remains unchanged.  The increase in gamma-ray flux produced by lensing, however, would result
in an underestimated burst distance ($z\propto$ luminosity/flux).  This is problematic for
detecting inconsistent GRB and galaxy redshifts because lensing will move the
pseudo-redshift of the GRB toward the redshift of the foreground galaxy.

Treating the lensing galaxy as a single isothermal sphere, the strongest lensing
effect would be produced by a low-redshift galaxy lying directly along the line-of-sight
to a high-redshift GRB (angular offset between source and lens $\lesssim0\farcs3$).
Random chance, however, favors less exact alignments.  At greater angular separations ($>0\farcs3$, where magnification only depends weakly
on angular separation),
only very large galaxies ($>$L$_*$ or $\sigma>100$km/s) can produce more than a factor of 2 in magnification of the
source. Consider a scenario in which a 0.1L$_*$ galaxy at z=0.3 lenses a GRB at z=3, with an angular
separation of $0\farcs5$.  This produces only a 15\% increase in brightness.
Lensing, therefore, only becomes a concern when a burst aligns very closely with a $>$L$_*$ galaxy.
However, there is only a $~10^{-4}$ chance that the line-of-sight to a GRB at
redshift $z=5$ would pass within $0\farcs3$ of the gravitational center of
a $>$L$_*$ galaxy.  For bursts that occur at lower redshift, this probability
only decreases.  Hence, while lensing will always occur with chance alignments, the effects
will generally be of little consequence.

Currently, there are only a few situations in which any kind of mis-match between a GRB and proposed host is likely
to be noted.  One example of this would be when no SN is detected in a low-redshift galaxy associated with a long-duration GRB.
A second example would be the association of a long-duration burst with an elliptical galaxy with little to no ongoing star-formation.
In either event, it is difficult to make a firm conclusion because of the possibility
that not all GRBs are produced in the canonical fashion.  One can question the cause of a GRB as easily as question its redshift.

To avoid contaminating host galaxy samples with false-hosts, it would be useful to have a
way to separate true and false hosts.  Unfortunately, no exact method of doing so exists.
Some strategies are:
\begin{itemize}
\item \textbf{Gamma-ray luminosity indicators:}
        While luminosity indicators \citep[e.g.][]{Amati+02,Ghirlanda+04,Liang+Zhang05} cannot currently
	produce accurate redshifts for individual bursts, they may eventually be able to
	be used in such a manner.
	In that case, however, a discrepancy could either indicate a chance superposition
        or a very unusual GRB, so that interpretation of such a mismatch
        is unclear. Lensing effects could also invalidate this method.

\item \textbf{Detection (or non-detection) of SNe:}
	This is only meaningful if long-duration bursts are all formed in core-collapse SNe
        and are similar in brightness to GRB 980425/SN 1998bw, which is assumed to be the archetypal GRB-related SN.  
	This is further limited to GRBs that occur at relatively low redshifts ($z\lesssim0.7$).
        Host-galaxy extinction is also a complicating factor that may 
	obscure an underlying SN.

\item \textbf{Optical afterglow spectral absorption features:}
	A GRB must occur behind the absorbing material, so line
	detection actually only gives a lower limit on redshift (though
	this is generally taken to be the GRB's redshift).
	Redshifts derived in this manner can be compared to
	redshifts measurements of the proposed host galaxy after
	the afterglow has faded. Only a very bright afterglow
	could possibly show absorption features independent
	of the superimposed galaxy, however.  The exception is high redshift bursts whose
	redshifts can be accurately determined by detection of the Lyman-$\alpha$ break.
	Alternatively,
	an afterglow with no absorption features might
	be assumed to be at low redshift because
	of the complete lack of absorption features
	along the line-of-sight.

\item \textbf{Visual detection of true host galaxy:}
	In deep images, the true host galaxy may be detectable
	through the foreground galaxy.  An example of this situation is
	GRB 060912a which occurred in a $z=0.937$ galaxy that
	overlaps with the outskirts of a large $z=0.0936$ galaxy \citep{Levan+07}.

\item \textbf{X-ray afterglow spectral emission lines:}  
	Such emission lines have been suggested to be produced by material associated 
	with the GRB (outflows, disks, etc.) so they could serve as accurate
	redshift indicators.  This is very unproven method, however,
	because line identification is inexact, reported lines
	have been at only low significance, and few GRB X-ray
	spectra contain these lines \citep[e.g.][]{Sako+05}.

\item \textbf{Strong lensing of GRB afterglows:} 
	This would be extremely unlikely for a single foreground galaxy,
	(as shown in Appendix A) but could potentially be caused by a foreground galaxy cluster.  
\end{itemize}

\section{Conclusions}
We have studied the galaxy population surrounding a sample of \totalnum\ GRBs.
Typically 1\% of the sky near the positions of GRBs is covered
by galaxies with $I\leq21.5$. 
With $\sim125$ \textit{Swift} GRBs with detected optical afterglows, the probability
that no chance alignments have been detected is $(0.99)^{125} = 28$\%.  Indeed,
approximately 1 superposition between the GRB and a foreground galaxy
is expected.  While it is possible that no such chance alignments have yet been observed, as \textit{Swift} detects
more and more GRBs with optical afterglows the likelihood of such an event only increases.
While most GRB/galaxy associations noted in the literature are almost certainly
correct, caution is required when making sweeping conclusions from only one or two
GRB/host galaxy associations.

We have also considered galaxies associated with X-ray afterglows.  
Over 250 X-ray afterglows, with typical error radii of $\sim2''$, have been detected by \textit{Swift}.
These numbers guarantee that some galaxies associated with X-ray afterglow will be falsely identified as hosts.
In fact, approximately half of the 14 X-ray afterglow error region/galaxy coincidences in our sample may exist only by chance.  
Even with large samples, therefore, using X-ray afterglow-identified host galaxies to draw conclusions about GRBs causes confusion.

\acknowledgments
We thank SMARTS observers D. Gonzalez, J. Espinoza and A. Pasten for their dedication
and S. Tourtellotte for assistance with optical data reduction. 
We are also grateful to P. G. van Dokkum and P. Natarajan for 
useful discussions about galaxies and lensing.
This work is supported by NSF Graduate Fellowship DGE0202738 
to BEC and NSF/AST grants 0407063 and 0707627 and \textit{Swift}
grant NNG05GM63G to CDB.

\clearpage
\begin{deluxetable}{lrrccccclll}
\tabletypesize{\scriptsize}
\tablecolumns{11}
\tablewidth{0pc}
\tablecaption{Properties of \totalnum\ SMARTS $I$-band GRB fields}
\tablehead{
	 \colhead{}                       		 	   & 
	 \multicolumn{3}{c}{X-ray AG Coordinates and Radius\tablenotemark{a}} & 
         \colhead{Optical}                          		   & 
         \colhead{Limiting}                          		   &
         \multicolumn{2}{c}{Overlap Probability}                  &
	\colhead{GRB}			&
	 \colhead{Galaxy/AG Overlaps}	\\	
         \colhead{GRB}                          &		
         \colhead{RA}                          &		
         \colhead{DEC}                          &
         \colhead{('')}                          &
         \colhead{AG?}                          &
         \colhead{$I$ mag\tablenotemark{b}}                          &
         \colhead{Optical\tablenotemark{c}}                          &
         \colhead{X-ray}		 	      &
	\colhead{redshift\tablenotemark{f}}	&
         \colhead{(and $I$ mag of galaxy)}          
}
\startdata
050128	&	14:38:17.66 &	-34:45:54.7	&	2.7	&	...	&	22.4	&	0.012	&	0.118&...&... \\
050219a	&	11:05:39.13 &	-40:41:02.6	&	3.5	&	...	&	22.5	&	0.010	&	0.154&...& X-ray (18.9)\\
050223	&	18:05:32.66 &	-62:28:20.4	&	3.2	&	...	&	22.3	&	0.017	&	0.194&0.584\tablenotemark{g}& X-ray (21.1)\\
050315	&	20:25:54.21 &	-42:36:02.5	&	1.4	&	yes	&	22.5	&	0.018	&	0.078&1.949&...	\\
050318	&	03:18:51.04 &	-46:23:43.5	&	2.7	&	yes	&	22.4	&	0.010	&	0.088&1.44&...	\\
050401	&	16:31:28.84 &	+02:11:14.5	&	1.8	&	yes	&	22.6	&	0.009	&	0.088&2.9&...	\\
050408	&	12:02:17.36 &	+10:51:10.5	&	1.2	&	yes	&	22.4	&	0.011	&	0.048&1.236&...	\\
050412	&	12:04:25.18 &	-01:12:00.8	&	6.9	&	...	&	22.6	&	0.011	&	0.322&...& X-ray (21.7)\\
050416a	&	12:33:54.57 &	+21:03:26.9	&	0.6	&	yes	&	22.7	&	0.012	&	0.036&0.6535\tablenotemark{g}& X-ray+optical (22.2)\\
050502b	&	09:30:10.06 &	+16:59:46.5	&	1.0	&	yes	&	22.3	&	0.006	&	0.014&...&...	\\
050509b\tablenotemark{d} & 	12:36:14.06 &	+28:59:07.2	&	3.4	&	...	&	22.3	&	0.023	&	0.190&0.226\tablenotemark{g}&...	\\
050520	&	12:50:05.94 &	+30:27:03.2	&	2.4	&	...	&	22.1	&	0.009	&	0.038&...&...	\\
050525	&	18:32:32.67 &	+26:20:21.2	&	2.1	&	yes	&	21.7	&	0.009	&	0.044&0.606\tablenotemark{g}&...	\\
050714b	&	11:18:47.75 &	-15:32:49.3	&	2.1	&	...	&	22.0	&	0.008	&	0.062&...& X-ray (20.9)\\
050724\tablenotemark{d}	&	16:24:44.32 &	-27:32:26.4	&	5.0	&	yes	&	21.6	&	0.035	&	0.572&0.258\tablenotemark{g}& X-ray+optical (17.4)\\
050726	&	13:20:12.16 &	-32:03:51.0	&	3.7	&	...	&	22.8\tablenotemark{e}	&	0.025	&	0.208&...&...	\\
050730	&	14:08:17.12 &	-03:46:16.3	&	1.4	&	yes	&	22.0	&	0.005	&	0.024&3.967&...	\\
050801	&	13:36:35.51 &	-21:55:42.7	&	5.0	&	yes	&	22.1\tablenotemark{e}	&	0.011	&	0.196&1.56&...	\\
050803	&	23:22:37.87 &	+05:47:09.8	&	1.8	&	...	&	22.5	&	0.019	&	0.090&...&...	\\
050822	&	03:24:27.22 &	-46:02:00.0	&	0.7	&	...	&	22.9	&	0.018	&	0.034&...&...	\\
050826	&	05:51:01.69 &	-02:38:37.6	&	2.6	&	yes	&	21.5	&	0.005	&	0.056&0.296\tablenotemark{g}& X-ray+optical (19.5)\\
050827	&	04:17:09.58 &	+18:12:00.4	&	1.3	&	...	&	21.3	&	0.005	&	0.018&...&...	\\
050915a	&	05:26:44.86 &	-28:00:59.9	&	1.4	&	yes	&	23.0	&	0.015	&	0.060&...&...	\\
050922b	&	00:23:13.19 &	-05:36:15.6	&	2.7	&	...	&	22.5	&	0.008	&	0.074&...&...	\\
051001	&	23:23:48.73 &	-31:31:23.3	&	1.5	&	...	&	23.1	&	0.014	&	0.050&...&...	\\
051006	&	07:23:14.03 &	+09:30:21.9	&	4.3	&	yes	&	22.5	&	0.028	&	0.292&...&...	\\
051021a	&	01:56:36.41 &	+09:04:04.3	&	2.1	&	yes	&	22.8	&	0.007	&	0.072&...&...	\\
051117b	&	05:40:43.21 &	-19:16:27.2	&	2.0	&	...	&	22.8	&	0.010	&	0.078&...& X-ray (20.3)\\
051210\tablenotemark{d}	& 	22:00:41.26 &	-57:36:46.5	&	2.9	&	...	&	21.9	&	0.009	&	0.108&...&...	\\
060108	&	09:48:02.00 &	+31:55:08.0	&	1.3	&	yes	&	22.4	&	0.010	&	0.046&$<3.2$&...	\\
060115	&	03:36:08.18 &	+17:20:44.3	&	2.6	&	yes	&	22.1	&	0.005	&	0.046&3.53&...	\\
060116	&	05:38:46.14 &	-05:26:15.2	&	3.1	&	yes	&	22.3	&	0.002	&	0.028&6.6\tablenotemark{h}&...	\\
060204b	&	14:07:15.05 &	+27:40:36.9	&	1.6	&	yes	&	22.4	&	0.024	&	0.068&...&...	\\
060211b	&	05:00:17.12 &	+14:56:58.1	&	1.9	&	...	&	21.3	&	0.005	&	0.040&...&...	\\
060213	&	09:26:23.05 &	-09:06:43.4	&	3.2	&	...	&	23.1	&	0.021	&	0.186&...&...	\\
060313\tablenotemark{d}	&	04:26:28.41 &	-10:50:40.7	&	2.4	&	yes	&	22.6	&	0.008	&	0.066&$<1.7$&...	\\
060505	&	22:07:03.39 &	-27:48:53.0	&	2.5	&	yes	&	22.7	&	0.010	&	0.104&0.089\tablenotemark{g}& X-ray+optical (17.5), X-ray (20.2)\\
060526	&	15:31:18.23 &	+00:17:05.4	&	1.6	&	yes	&	22.1	&	0.011	&	0.044&3.21&...	\\
060602a	&	09:58:16.93 &	+00:18:12.8	&	2.3	&	yes	&	21.9	&	0.004	&	0.018&0.787\tablenotemark{g}&...	\\
060604	&	22:28:55.01 &	-10:54:55.9	&	1.0	&	yes	&	21.2\tablenotemark{e}	&	0.003	&	0.008&2.68&...	\\
060614	&	21:23:32.04 &	-53:01:37.0	&	1.4	&	yes	&	23.3	&	0.022	&	0.076&0.125\tablenotemark{g}& X-ray+optical (21.9)\\
060708	&	00:31:13.79 &	-33:45:33.3	&	1.4	&	yes	&	22.5	&	0.011	&	0.056&$<2.3$&...	\\
060714	&	15:11:26.43 &	-06:33:59.5	&	1.8	&	yes	&	22.2	&	0.009	&	0.062&2.71&... 	\\
060719	&	01:13:43.69 &	-48:22:51.0	&	1.5	&	yes	&	22.9	&	0.010	&	0.066&...&...	\\
060729	&	06:21:31.90 &	-62:22:12.6	&	0.5	&	yes	&	22.3	&	0.009	&	0.020&0.54&...	\\
060801\tablenotemark{d}	&	14:12:01.35 &	+16:58:53.7	&	2.4	&	...	&	22.1	&	0.005	&	0.036&...&...	\\
060805	&	14:43:43.46 &	+12:35:11.8	&	1.8	&	...	&	22.1	&	0.009	&	0.050&...&...	\\
060814	&	14:45:21.29 &	+20:35:10.7	&	0.9	&	yes	&	21.6	&	0.004	&	0.008&0.84\tablenotemark{g}&...	\\
060904b	&	03:52:50.45 &	-00:43:30.0	&	1.7	&	yes	&	22.2	&	0.018	&	0.080&0.703&...	\\
060912a	&	00:21:08.11 &	+20:58:18.9	&	1.9	&	yes	&	22.5	&	0.017	&	0.088&0.937\tablenotemark{g}&...	\\
060919	&	18:27:41.89 &	-51:00:51.1	&	2.6	&	...	&	22.0	&	0.022	&	0.222&...& X-ray (21.1)\\
060923c	&	23:04:28.36 &	+03:55:28.4	&	1.8	&	yes	&	21.8\tablenotemark{e}	&	0.013	&	0.078&...&...	\\
060926	&	17:35:43.86 &	+13:02:19.0	&	4.6	&	yes	&	21.2	&	0.016	&	0.250&3.208&...	\\
060927	&	21:58:12.23 &	+05:21:52.2	&	3.8	&	yes	&	22.8	&	0.012	&	0.170&5.6& X-ray (22.0)\\
061004	&	06:31:10.93 &	-45:54:23.8	&	1.6	&	...	&	22.2	&	0.011	&	0.068&...&...	\\
061006	&	07:24:07.56 &	-79:11:56.6	&	2.4	&	yes	&	22.0	&	0.005	&	0.056&...&...	\\
061007	&	03:05:19.55 &	-50:30:01.8	&	1.5	&	yes	&	22.4	&	0.012	&	0.042&1.261&...	\\
061021	&	09:40:36.04 &	-21:57:04.9	&	1.3	&	yes	&	22.4	&	0.010	&	0.028&$<2.0$& X-ray+optical (21.4)\\
061025	&	20:03:37.18 &	-48:14:28.9	&	3.4	&	yes	&	22.0	&	0.011	&	0.144&...&...	\\
061102	&	09:53:37.84 &	-17:01:26.0	&	2.9	&	...	&	22.4	&	0.010	&	0.100&...&...	\\
061110a	&	22:25:09.92 &	-02:15:31.6	&	1.6	&	yes	&	22.4	&	0.020	&	0.084&0.758\tablenotemark{g}&... 	\\
061121	&	09:48:54.50 &	-13:11:43.0	&	1.0	&	yes	&	22.6	&	0.008	&	0.026&1.314& X-ray+optical (22.2)\\
061201\tablenotemark{d}	&	22:08:32.09 &	-74:34:48.2	&	3.6	&	yes	&	22.1	&	0.006	&	0.104&...&...	\\
061202	&	07:02:05.54 &	-74:41:54.6	&	1.5	&	...	&	22.2	&	0.008	&	0.052&...&...	\\
061217\tablenotemark{d}	&	10:41:39.32 &	-21:07:22.1	&	3.8	&	...	&	22.5	&	0.009	&	0.142&0.827\tablenotemark{g}&...	\\
070224	&	11:56:06.57 &	-13:19:49.4	&	1.7	&	yes	&	22.7	&	0.016	&	0.076&...&...	\\
070306	&	09:52:23.32 &	+10:28:55.4	&	1.1	&	yes	&	22.4	&	0.008	&	0.032&1.497\tablenotemark{g}&...	\\
070318	&	03:13:56.79 &	-42:56:47.7	&	1.4	&	yes	&	21.8	&	0.003	&	0.002&0.840\tablenotemark{g}&...	\\
070330	&	17:58:10.53 &	-63:47:35.0	&	2.4	&	yes	&	22.2	&	0.018	&	0.136&...&...	\\
070419b	&	21:02:49.91 &	-31:15:47.2	&	1.7	&	yes	&	22.5	&	0.021	&	0.096&...&...	\\
070429a	&	19:50:48.71 &	-32:24:15.1	&	1.9	&	...	&	22.2	&	0.020	&	0.126&...&...	\\
070508	&	20:51:12.10 &	-78:23:03.4	&	2.1	&	yes	&	22.2	&	0.008	&	0.070&0.82\tablenotemark{g}&...	\\
\enddata
\tablenotetext{a}{X-ray afterglow coordinates and 90\% confidence error radii for all bursts are taken from \cite{Butler07}.}
\tablenotetext{b}{3$\sigma$ limiting $I$ magnitude for point sources in the field, corrected for Galactic reddening.  
	Unless otherwise indicated, photometric calibration is typically accurate to 0.05 magnitudes.}
\tablenotetext{c}{Based on the ISOAREAF\_IMAGE area of each galaxy.}
\tablenotetext{d}{Short-duration burst.}
\tablenotetext{e}{Photometrically calibrated using USNO-B1.0 I2 magnitudes, typically accurate to 0.2 magnitudes.}
\tablenotetext{f}{Redshift references: 050223--Pellizza et al. 2006; 050315--Kelson \& Berger 2005;   
050318--Berger \& Mulchaey 2005; 050401--Fynbo et al. 2005; 050408--Berger et al. 2005;
050416a--Cenko et al. 2005; 050509b--Prochaska et al. 2005a; 050525--Foley et al. 2005;
050724--Prochaska et al. 2005b; 050730--Chen et al. 2005; 050801--de Pasquale et al. 2007;
050826--Halpern \& Mirabal 2006; 060108--Oates et al. 2006; 060115--Piranomonte et al. 2006a;
060116--Piranomonte et al. 2006b; 060313--Roming et al. 2006; 060505--Ofek et al. 2006;
060526--Berger \& Gladders 2006; 060602a--Jakobsson et al. 2007a; 060604--Castro-Tirado et al. 2006;
060614--Price et al. 2006; 060708--Jakobsson et al. 2006a; 060714--Jakobsson et al. 2006b;
060729--Thoene et al. 2006a; 060814--Thoene et al. 2007; 060904b--Fugazza et al. 2006;
060912a--Jakobsson et al. 2006c; 060926--D'Elia et al. 2006; 060927--Fynbo et al. 2006;
061007--Osip et al. 2006; 061021--Thoene et al. 2006b; 061110a--Fynbo et al. 2007;
061121--Bloom et al. 2006b; 061217--Berger 2006; 070306--Jaunsen et al. 2007;
070318--Chen et al. 2007; 070508--Jakobsson et al. 2007b
}
\tablenotetext{g}{Redshift derived from galaxy emission, not measured from afterglow absorption.}
\tablenotetext{h}{Possibly \textit{z}=3.8-4.5, depending on an uncertain amount of extinction.}
\end{deluxetable}

\clearpage

\begin{deluxetable}{lllllll}
\tabletypesize{\small}
\tablecolumns{7}
\tablewidth{0pc}
\tablecaption{Optical afterglow chance coincidence probabilities}
\tablehead{
         \colhead{}                         &
         \colhead{Mag.}                    &
         \colhead{}                         &
         \colhead{Median}                   &
         \colhead{\# expected}      &
         \colhead{\# observed}       &
         \colhead{Prob. of observed} \\
         \colhead{Galaxy area}                         &
         \colhead{cutoff}                    &
         \colhead{Sample\tablenotemark{a}}          &
         \colhead{prob.\tablenotemark{b}}                   &
         \colhead{coincidences}      &
         \colhead{coincidences}       &
         \colhead{coincidences\tablenotemark{c}} 
}
\startdata
ISOAREAF\_IMAGE      &  none    &       total& \Imedianarea    &       \Iexpected      &       7&      0.004\%         \\
                     &  none    &       OAG  & \ImedianOAG     &       \IexpectedOAG   &       7&      0.0002\%        \\
                     &  $I\leq21.5$ &   total& \Imedianareawcut&       \Iexpectedwcut   &      4&      0.98\%          \\
                     &  $I\leq21.5$ &   OAG  &  \ImedianOAGwcut &       \IexpectedOAGwcut&      4&      0.18\%          \\
Ellipse ($v=3$)      &  none    &       total & \Emedianarea    &       \Eexpected      &       7&      0.04\%          \\
                     &  none    &       OAG  & \EmedianareaOAG &       \EexpectedOAG   &       7&      0.002\%         \\
                     &  $I\leq21.5$ &   total & \Emedianareawcut&       \Eexpectedwcut  &       4&      3.2\%           \\
                     &  $I\leq21.5$ &   OAG  & \EmedianareawcutOAG&    \EexpectedwcutOAG&      4&      0.68\%          \\
\enddata
\tablenotetext{a}{The sample of GRB fields used.  ``Total'' refers to all 72 GRBs.  ``OAG'' refers to the 47 GRBs
for which optical afterglows were detected.}
\tablenotetext{b}{The median probability of a chance coincidence between an optical afterglow and a galaxy.}
\tablenotetext{c}{Probability of having observed the given number of optical afterglow/galaxy coincidences
at random given the expected number of random coincidences, calculated using Monte Carlo simulations.}
\end{deluxetable}

\clearpage

\begin{figure}
\includegraphics[width=1\textwidth]{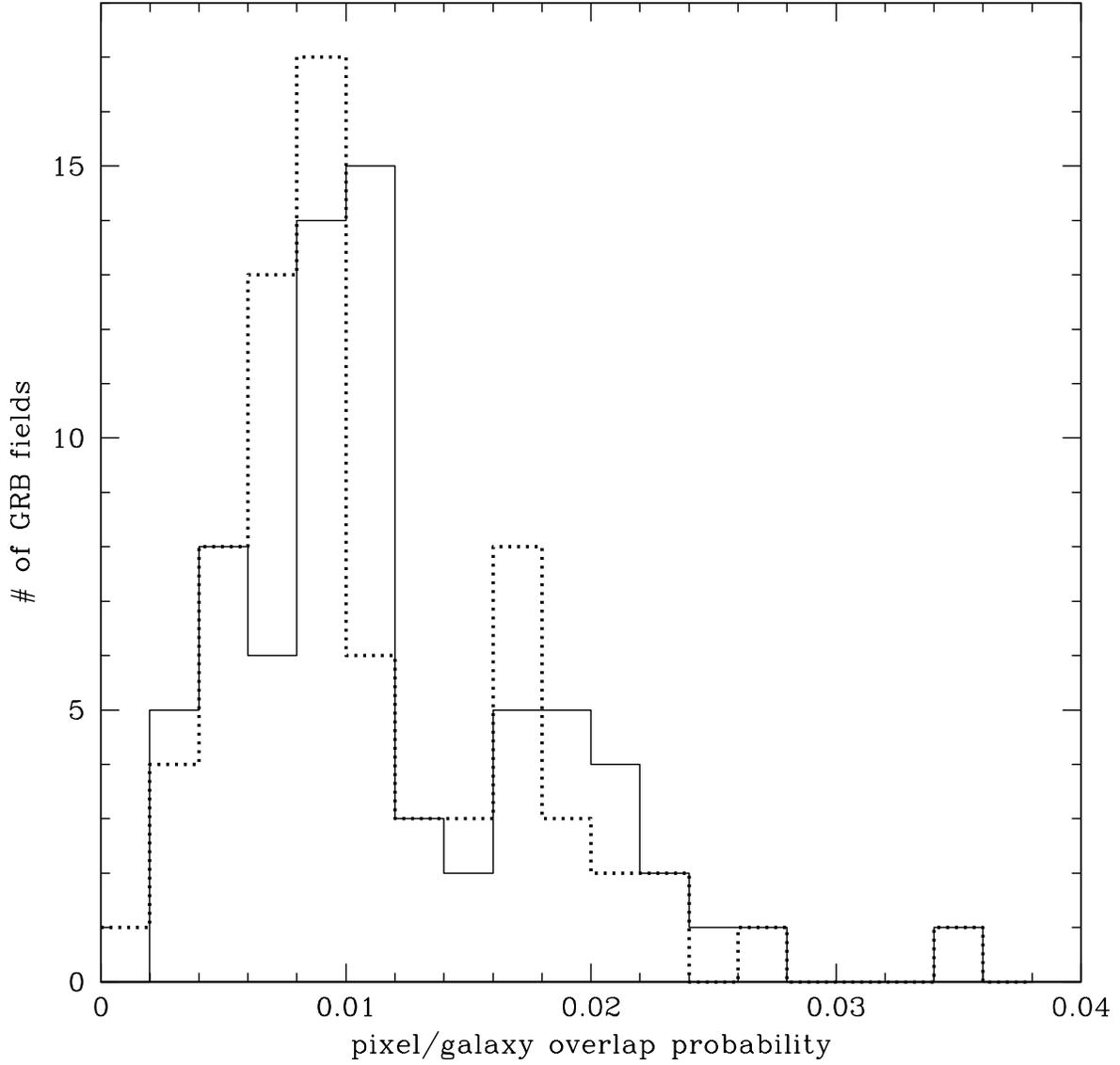}
\caption{Histogram of the probability of any pixel in a GRB field falling
inside of a galaxy (\textit{solid line} - median of \Imedianarea) or inside a
galaxy with magnitude $I\leq21.5$ (\textit{dotted line} - median of \Imedianareawcut).
The area of each galaxy is given by the SExtractor output ISOAREAF\_IMAGE.
}
\end{figure}

\begin{figure}
\includegraphics[width=1\textwidth]{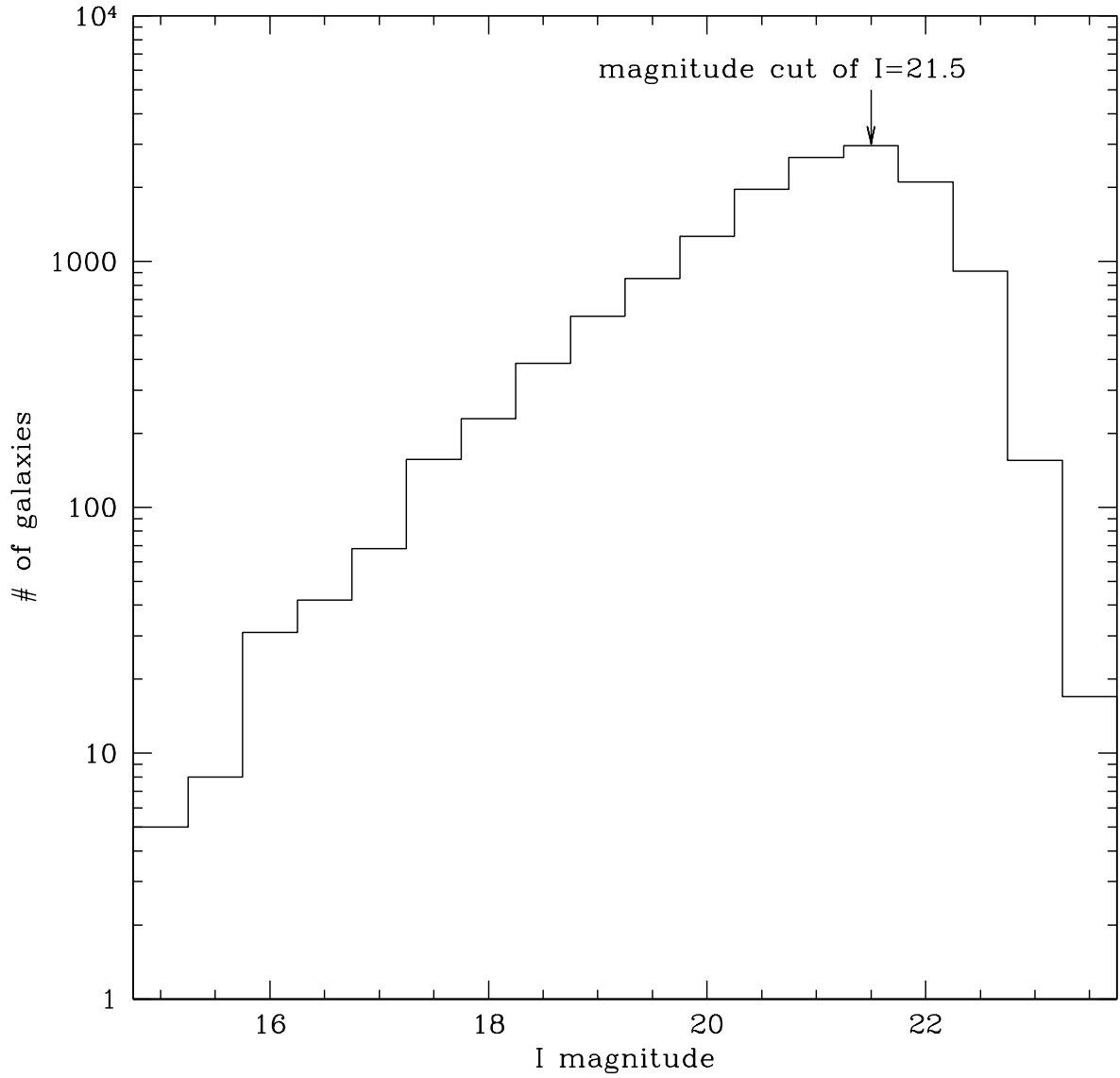}
\caption{Magnitude histogram of all galaxies in this sample.  The magnitude cut
of I=21.5 is chosen as the midvalue of the bin containing the maximum number
of galaxies.  The sample of dimmer galaxies is significantly affected by
incompleteness.
}
\end{figure}

\begin{figure}
\includegraphics[width=1\textwidth]{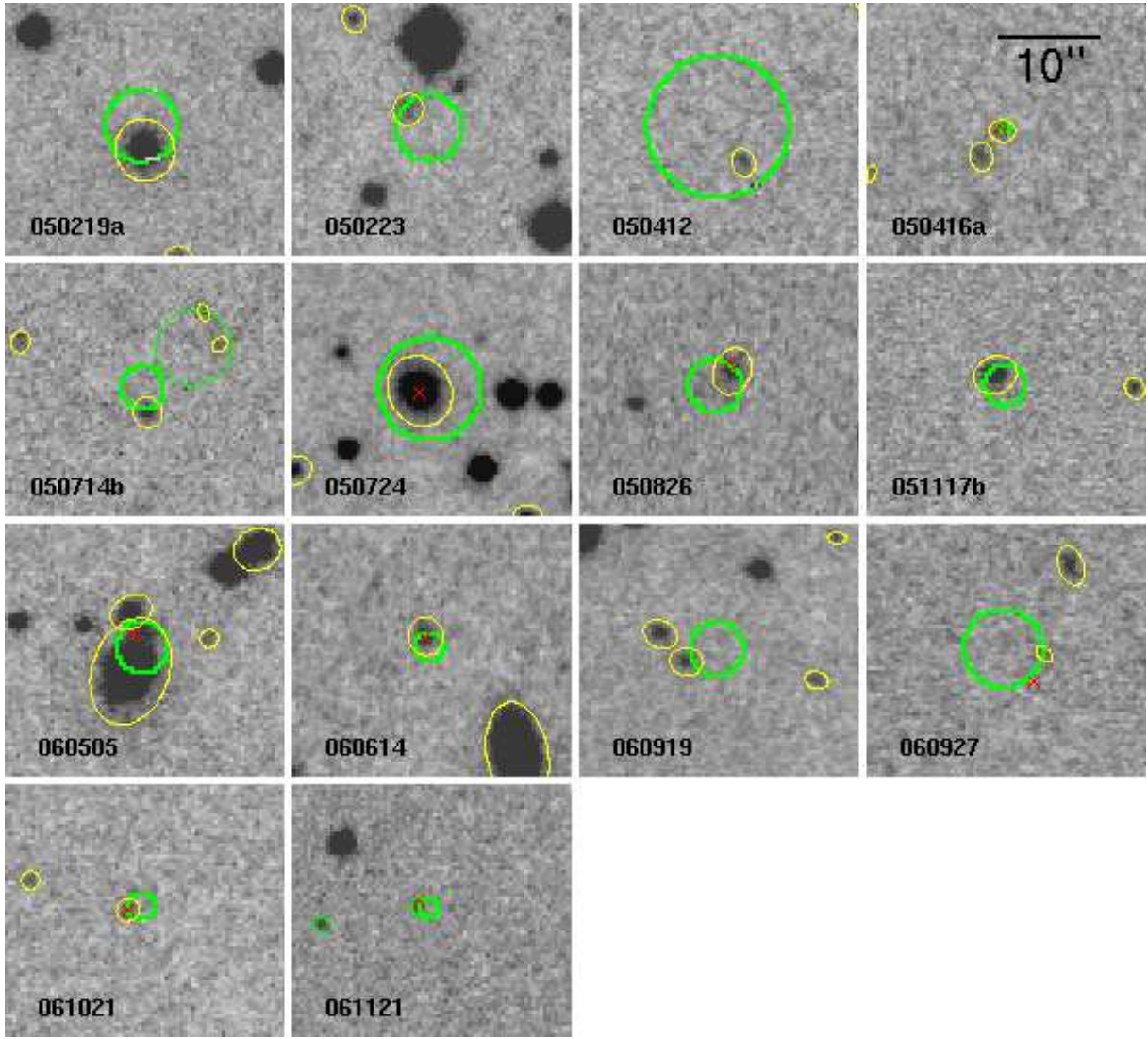}
\caption{Sections of each of the 14 fields in which the X-ray afterglow error region
overlaps a galaxy.  The X-ray afterglow error regions are shown as green circles.
SExtracted galaxies are indicated as yellow ellipses.  Each ellipse has major and minor axes
of 3$\times$A\_IMAGE and 3$\times$B\_IMAGE.  For bursts with detected optical afterglow,
the optical afterglow position is marked by a red x.
}
\end{figure}

\begin{figure}
\includegraphics[width=1\textwidth]{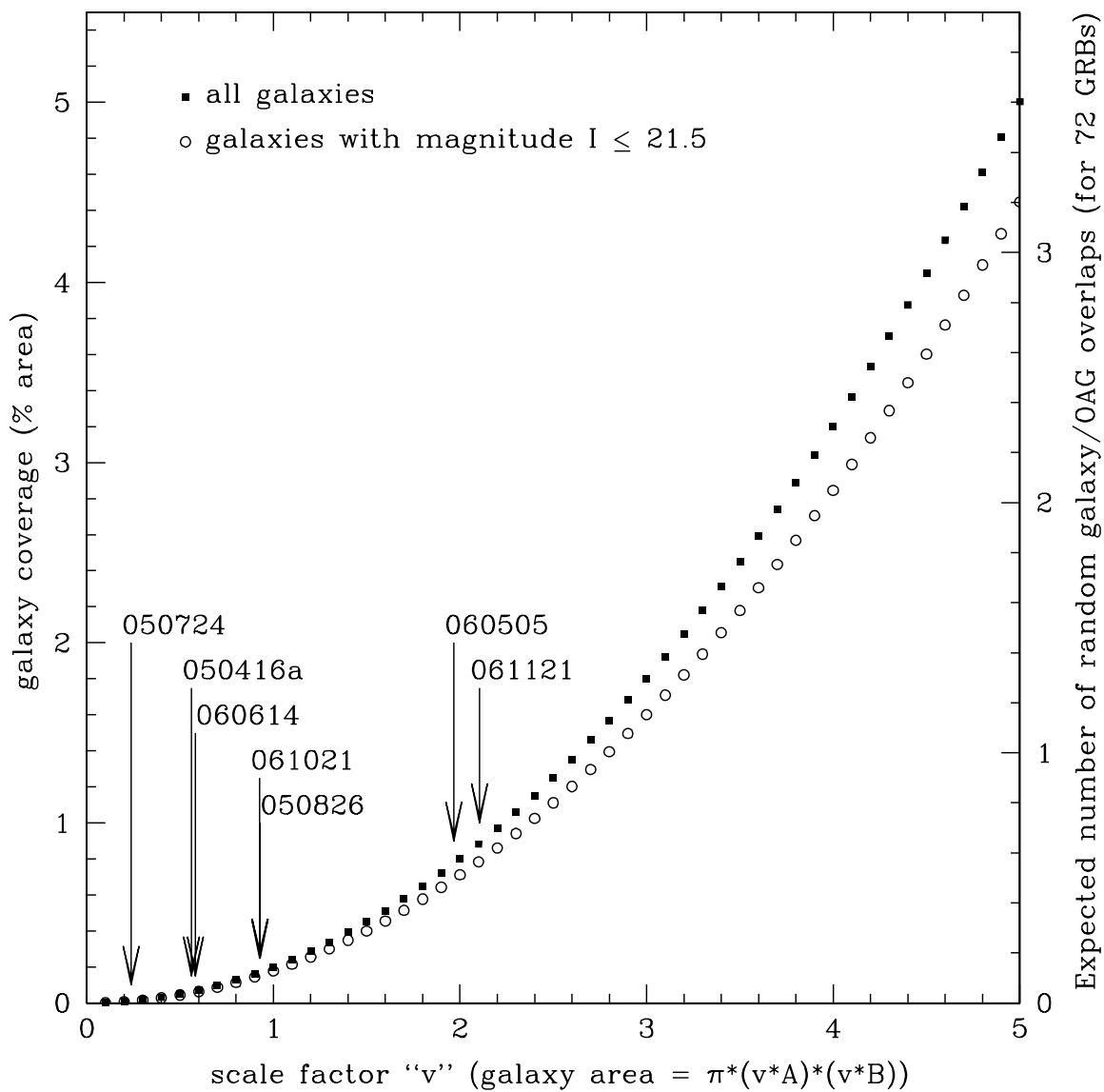}
\caption{
The probability of chance overlaps for \textit{(solid squares)} all galaxies or \textit{(open circles)} galaxies
with $I\leq21.5$ for increasing values of $v$, where $v$ is the scale factor used in defining galaxy size.
On the right is shown the corresponding expectation value for the 72 GRBs in this sample.
The $v$ values of the 7 GRBs with optical afterglow/galaxy coincidences are marked with arrows.
}
\end{figure}

\begin{figure}
\includegraphics[width=1\textwidth]{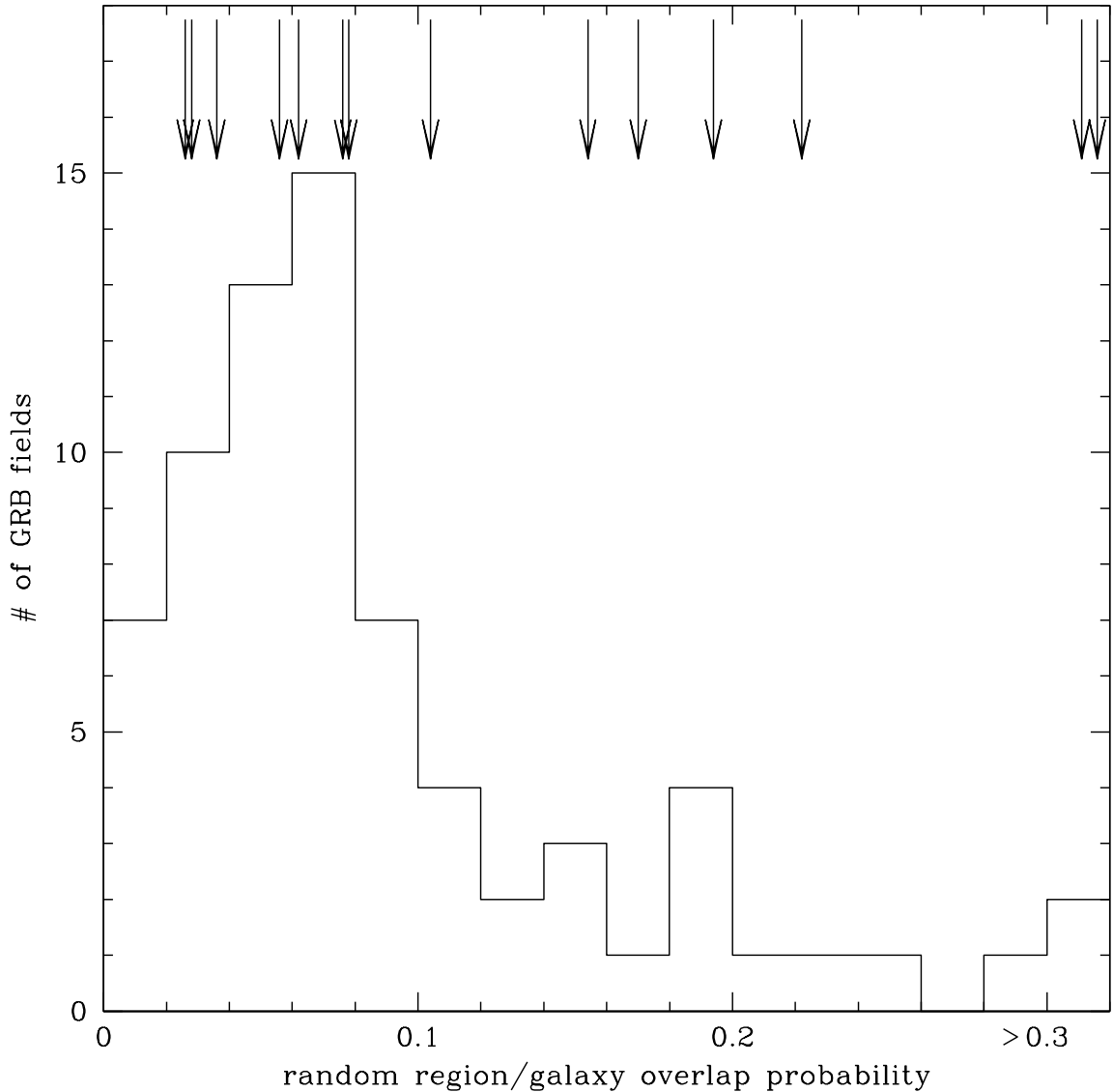}
\caption{Histogram of the probability of a random X-ray error region/galaxy overlap
in the \totalnum\ GRB fields examined.  The median probability is \Xmedianprob.  The
arrows indicate the actual probability of each field in which the X-ray
afterglow error region did coincide with one or more galaxies.  
The arrows over $>0.3$ indicate fields having actual probabilities of 32\% and 57\%.
A Kolmogorov-Smirnov (K-S) test on these data gives $P\sim0.4$, which indicates that
the distribution of probabilities in the 14 fields containing X-ray afterglow-associated galaxies
is not significantly different from the distribution of the entire sample.
}
\end{figure}

\begin{figure}
\includegraphics[width=1\textwidth]{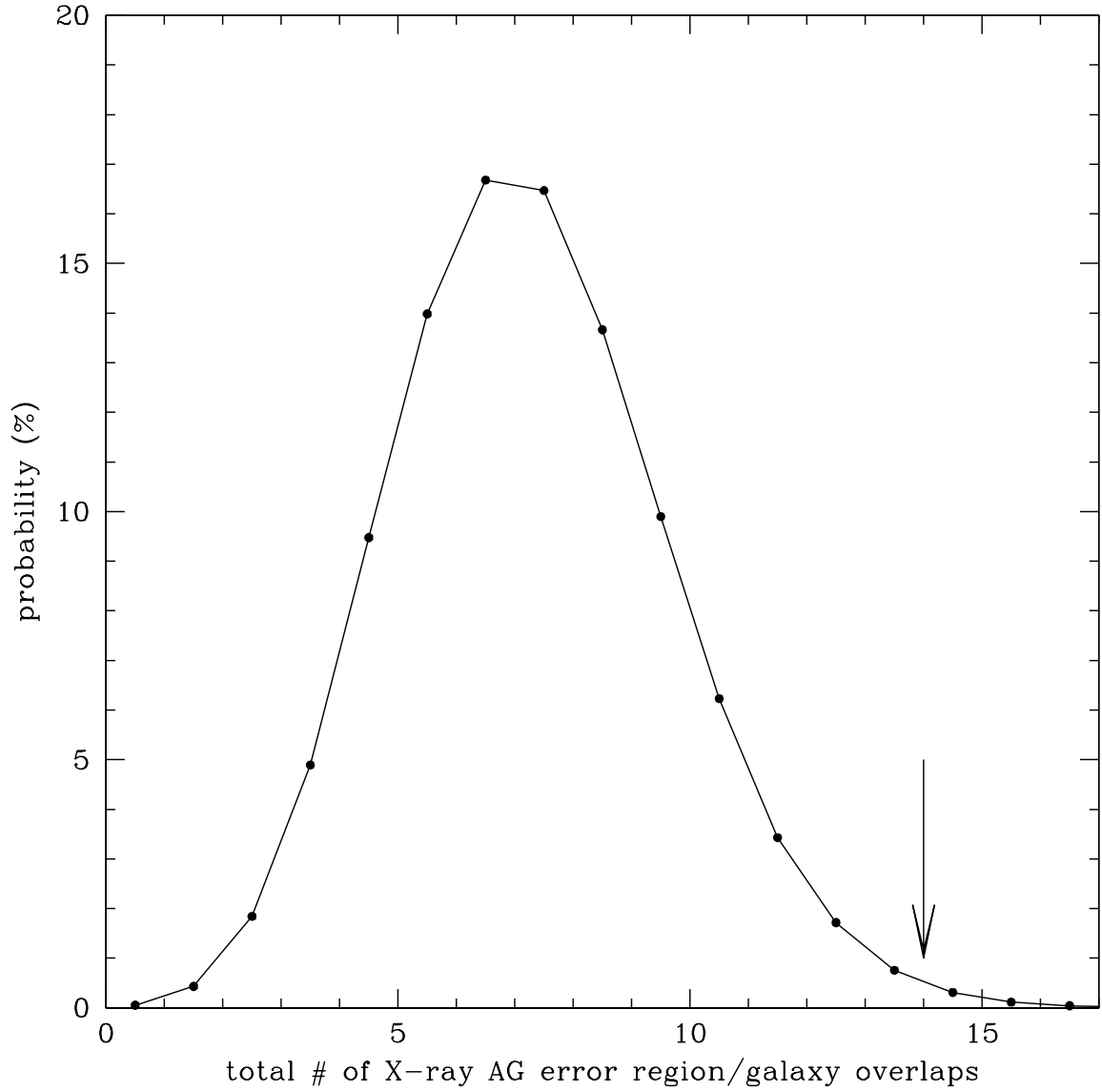}
\caption{Monte Carlo simulation of X-ray afterglow error region/galaxy coincidences within our \totalnum\ GRB fields.
Approximately 7 overlaps are expected, while \Xoverlapnum\ are observed (indicated
by the arrow).  The probability of observing \Xoverlapnum\ or more overlaps
is only \Xmontecarlo. 
}
\end{figure}

\begin{figure}
\includegraphics[width=1\textwidth]{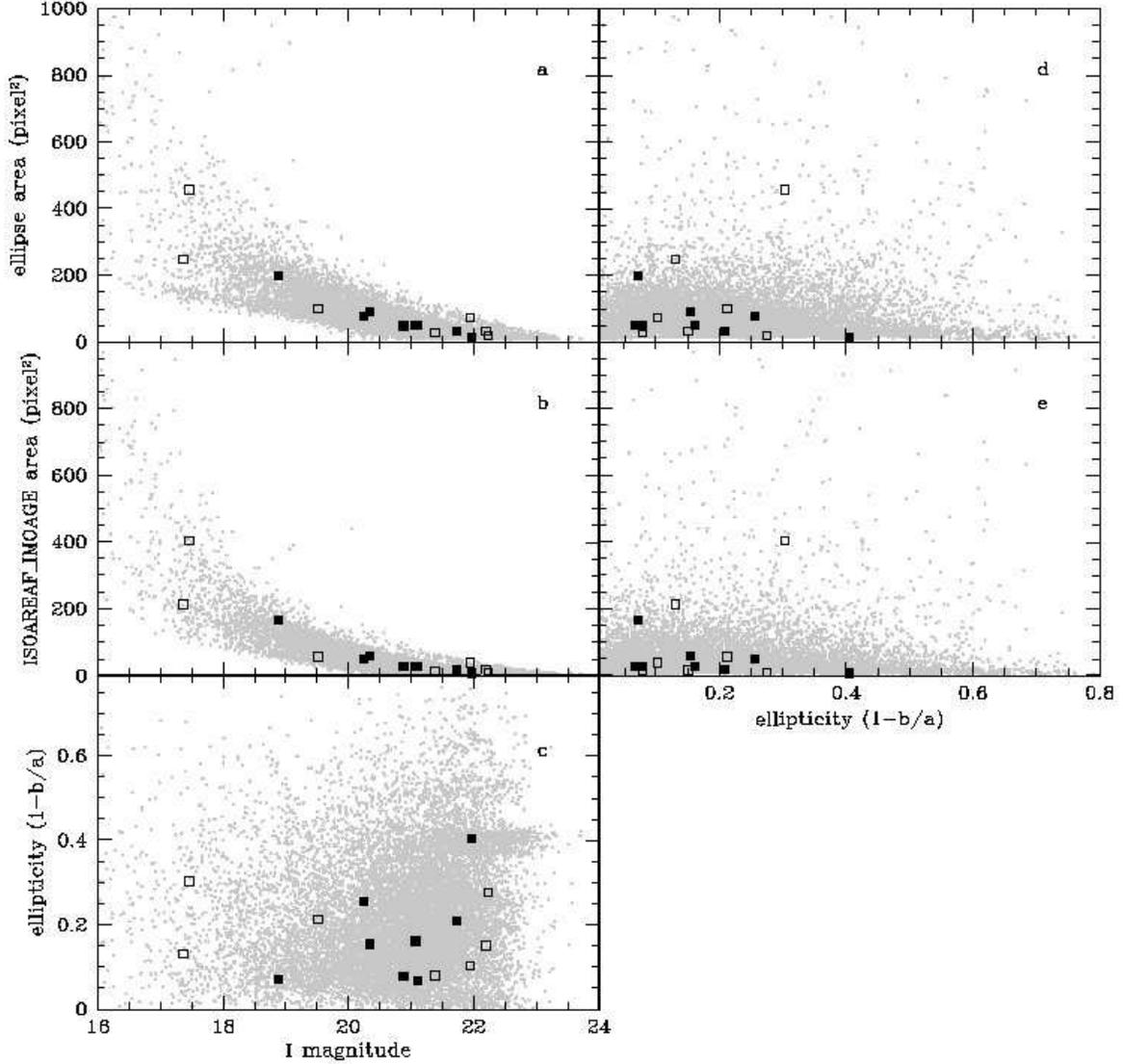}
\caption{Observed galaxy properties:
\textit{(a)} galaxy ellipse area versus magnitude, \textit{(b)} ISOAREAF\_IMAGE area versus magnitude,
\textit{(c)} ellipticity versus magnitude, \textit{(d)} galaxy ellipse area versus ellipticity
and \textit{(e)} galaxy ISOAREAF\_IMAGE area versus ellipticity.
The plot axes limits are chosen to emphasize the afterglow-associated galaxies and exclude some of the galaxies
in this sample. While not plotted, the galaxies are included in the Monte Carlo
simulations. The galaxies associated with
X-ray afterglows (\textit{filled and open squares}) are a typical
sample of all the field galaxies (\textit{gray points}).
The galaxies associated with both X-ray and optical afterglows (\textit{open squares})
differ from the field galaxy population at the $\sim2\sigma$ level
for all relationships involving magnitude (\textit{a,b,c}).}
\end{figure}

\end{document}